\documentclass[aps,pra,eqsecnum,twocolumn]{revtex4}
\usepackage{graphicx}
\usepackage{bm}
\usepackage{amsmath,amssymb,amsthm,dsfont,bm}
\usepackage{color}
\usepackage{wasysym}
\usepackage{physics}
\usepackage{ulem}
\usepackage{appendix}
\usepackage[dvipsnames]{xcolor}

\usepackage[english]{babel} 
\usepackage[utf8]{inputenc} 
\usepackage{hyperref}
\usepackage{cancel}

\begin{document}
\preprint{}
\title{Quantum tests via inequalities for joint statistics }
\author{Alfredo Luis}
\affiliation{Departamento de \'Optica, Facultad de Ciencias F\'{\i}sicas, Universidad Complutense, 28040 Madrid, Spain}

\date{\today}

\begin{abstract}
In this work we derive statistical inequalities whose violation is equivalent to the impossibility of describing the data by a genuine joint probability distribution. So they are witnesses of the failure of a common probability space. We examine whether the most significant quantum quasidistributions violate these inequalities. We examine also whether these inequalitites are violated by the joint distributions derived form a noisy joint measurement. As a relevant example we find violations for the maximally mixed state, as well as cases where all system states violate them. This points to the idea that  these results are more than a property of the system states, but are instead a property of the statistical structure of quantum mechanics. 
\end{abstract}
\maketitle

\section{Introduction}

Quantum theory is a statistical theory. It is therefore natural that the tests capable of revealing its deepest nature should also be of statistical character. The best example of this are Bell-like inequalities \cite{JB64,AF82,SGH21,SH22}.

In all these quantum tests, complementarity plays a fundamental role. Sooner or later quantumness turns out to be linked to the absence of joint probability distributions for incompatible observables \cite{AF82,MAL20,AR15,MA84,AA24,DGG05,AP00,AK00,HP04,AM08,TN11,AK14,JCh17}. It is worth noting that the idea of complementarity is also deeply rooted in classical theory \cite{GBAL20,MAL20}. As a first example we have the case of radiometry, with the impossibility, in principle, of determining the amount of light in a given spatial point propagating along a given direction. As a further example, we find a similar impossibility for defining the amount of light being polarized both linearly and circularly at the same time.

\bigskip

In this context there are tools that allow us to bridge the gap between quantum theory and classical statistics. This is the case of simultaneous, albeit noisy, measurements of incompatible observables, leading to a legitimate observable joint distribution \cite{MM90,WMM98,WMM93,WMM02,PB87,PB86,YLLO10,JM68,TH16}. In this regard, we have recently investigated Bell-like and complementarity tests based on this idea that allows us to apply all the tools of statistics. For example, we can examine outcomes of single measurements as well as the probability of test success as a function of the number of times the measurement is repeated \cite{MAL20,AL25a,AL25b,AL25c}. 

This approach is directly related to the families of quasi-probability distributions which are in principle well-founded descriptions of incompatible observables. They are powerful descriptions of quantum states alternative to the traditional Hilbert space techniques. Well-known examples include the Wigner function, the Glauber-Sudarshan function $P(\alpha)$, the Husimi Q-function, and the Kirkwood-Dirac distribution \cite{EW32,CG69,ES63,KH40,PD45,JK33,LBLHFG23,ABDDJLLLH24,CF11,LB26}. In fact, noisy measurements can be viewed precisely in this context as providing the joint distributions forbidden by the Hilbert-space formalism. This is well illustrated by Husimi’s Q-function \cite{AK65,CLG87,LP93a,LP93b,KL08}. 

\bigskip

In this work we derive a family of inequalities that should be satisfied by genuine joint probability distribution of pairs of dichotomic observables. Such inequalities can be regarded as suitable constraints that genuine joint distributions must comply. In other words, we derive explicit statistical inequalities whose violation is equivalent to the impossibility of describing the data by a genuine joint probability distribution. 

We examine whether the most significant quantum quasidistributions satisfy these inequalities. Moreover, we examine also whether these inequalitites are violated by joint distributions derived from a noisy joint measurement. As a relevant case we find violations for the maximally mixed state, as well as cases where all system states violate them. This points to the idea that  these results are more than a property of the system states, but are instead a property of the statistical structure of quantum mechanics. 

\section{Physical system} 

Our system is described by a two-dimensional Hilbert space. This is a qubit in any of its possible physical realizations. For definiteness, we may consider that the physical system in question is transverse polarization. In the classical domain this can be the polarization state of the electric field at a given spatial point, being represented by the $2 \times 2$ polarization matrix $\Gamma$. This can be as well the case of a two-mode single photon represented by a $2 \times 2$ density matrix $\rho$. 

Both $\rho$ or $\Gamma$ can be suitably expressed in terms of the Pauli matrices $\sigma_{1,2,3}$ and the $2 \times 2$ identity $\sigma_0 $ as
\begin{equation}
\label{GSp}
    \rho = \Gamma = \frac{1}{2} \left ( \sigma_0+ \bm{s} \cdot \bm{\sigma} \right ) ,
\end{equation}
where $\bm{s}$ is a real three-dimensional vector with $\bm{s}^2\leq 1$ and $\bm{\sigma}=(\sigma_1,\sigma_2,\sigma_3) $ is a vector with the three Pauli matrices.  For the sake of simplicity in the classical case units are chosen so that ${\rm tr} \Gamma = 1$. In classical optics $\bm{s}$ are actually the Stokes parameters, while for qubits they are Bloch vectors. 

We will consider three dichotomic observables $X=A,B,C$ with eigenvalues $x=a,b,c =\pm 1$. They can be described also by three-dimensional real unit vectors $\bm{s}_X$, so that the corresponding operators are of the form
\begin{equation}
    \hat{X}= \bm{s}_X \cdot \bm{\sigma},\quad \bm{s}^2_X =1 ,
\end{equation}
whose eigenvectors can be expressed as 
\begin{equation}
\label{eigen}
    |x\rangle \langle x | = \frac{1}{2} \left ( \sigma_0+ x \bm{s}_X \cdot \bm{\sigma} \right ) .
\end{equation}
Now that we are on the subject, we can provide some useful expressions for order-sensitive quantum correlations of observables. This is that after the Pauli-matrices relation
\begin{equation}
\label{oqc}
    \hat{X} \hat{Y}= \left ( \bm{s}_X \cdot \bm{s}_Y \right ) \sigma_0 + i \left ( \bm{s}_X \times \bm{s}_Y \right ) \cdot \bm{\sigma} .
\end{equation}
we have the quantum correlation
\begin{equation}
\label{mqc}
    \langle \hat{X} \hat{Y} \rangle = \bm{s}_X \cdot \bm{s}_Y+ i \left ( \bm{s}_X \times \bm{s}_Y \right ) \cdot \bm{s} .
\end{equation}

\section{Inequality} 

The statistical inequality we shall work with is 
\begin{equation}
\label{ine}
    p_{AB}(a,b) \leq p_{AC}(a,c) + p_{BC}(b,-c) ,
\end{equation}
where $p_{XY}(x,y)$ denotes the joint probability that variable $X$ takes value $x$ and variable $Y$ takes value $y$, always $x,y=\pm 1$. This is a Boole-type inequality following directly from the existence of an underlying joint probability distribution $p_{ACB}(a,c,b)$. It can be easily demonstrated from the equality 
\begin{equation}
p_{AB}(a,b) = p_{ABC}(a,b,c) +p_{ABC}(a,b,-c) ,
\end{equation}
together with 
\begin{eqnarray}
    & p_{ABC}(a,b,c) \leq  p_{AC}(a,c) , &\nonumber \\
    & & \\
& p_{ABC}(a,b,-c) \leq  p_{BC}(b,-c)  , & \nonumber
\end{eqnarray}
which follows from the nonnegativity of $p_{ACB}(a,c,b)$.

\bigskip

We can derive an alternative form for the inequality (\ref{ine}). To this end we consider the most general for the joint distribution of two dichotomic variables, 
\begin{equation}
\label{nvpXY}
    p_{XY} (x,y) = \frac{1}{4} \left ( 1 + x \langle X \rangle + y \langle Y \rangle+ xy \langle XY \rangle_p \right ) ,
\end{equation}
where $\langle XY \rangle_p $ means
\begin{equation}
  \langle XY \rangle_p = \langle YX \rangle_p = \sum_{x,y} xy  p_{XY} (x,y) ,
\end{equation}
where the subscript $p$ is to distinguish it from the quantum averages $\langle \hat{X}\hat{Y} \rangle $ or $\langle \hat{Y} \hat{X} \rangle$. This is relevant since the quantum theory does not properly define the correlation between two arbitrary observables. With this we have
\begin{equation}
\label{gf}
 p_{AC}(a,c) + p_{BC}(b,-c) = p_{AB}(a,b) + \eta ,
\end{equation}
where
\begin{equation}
\label{eta}
\eta =  \frac{1}{4} \left ( 1 - ab \langle AB \rangle_p +ac \langle AC \rangle_p - cb \langle CB \rangle_p \right ) ,
\end{equation}
so the inequality (\ref{ine}) is equivalent to $\eta \geq 0$. It is worth noting that the fulfillment of the statistical inequality rests solely on correlation terms that are largely undefined in the quantum case.

\bigskip

Actually we can show that the quantum theory is at odds with condition $\eta \geq 0$. To this we can consider the following quantum mean value
\begin{equation}
   \left \langle \left ( a \hat{A} -b \hat{B} + c \hat{C} \right )^2 \right \rangle,
\end{equation}
which is equal to 
\begin{equation}
   \left \langle \left ( a \hat{A} -b \hat{B} + c \hat{C} \right )^2 \right \rangle = \bm{S}^2  ,
\end{equation}
where
\begin{equation}
    \bm{S} = a \bm{s}_A - b \bm{s}_B + c \bm{s}_C .
\end{equation}
This leads to 
\begin{eqnarray}
& \left \langle \left ( a \hat{A} -b \hat{B} + c \hat{C} \right )^2 \right \rangle =  3 & \nonumber \\
& & \nonumber \\
&+ 2 \left (  - ab \langle \hat{A}\hat{B} \rangle_s +ac \langle \hat{A}\hat{C} \rangle_s - cb \langle \hat{B} \hat{C} \rangle_s \right ) , &
\end{eqnarray}
where $\langle \hat{A}\hat{B} \rangle_s$ denotes the corresponding symmetric quantum correlation
\begin{equation}
\label{soc}
\langle \hat{X} \hat{Y} \rangle_s = \frac{1}{2} \left ( \langle \hat{X} \hat{Y} \rangle + \langle \hat{Y} \hat{X} \rangle \right ) = \bm{S}_X \cdot \bm{S}_Y.
\end{equation}
Therefore if in Eq. (\ref{eta}) we replace $\langle XY \rangle_p$ by $\langle \hat{X} \hat{Y} \rangle_s$ as a proper quantum translation we get
\begin{equation}
\label{eqc}
    \eta = \frac{1}{8} \left ( \bm{S}^2 -1\right ) ,
\end{equation}
and then
\begin{equation}
\eta \geq 0 \longleftrightarrow \bm{S}^2 \geq 1 .
\end{equation}
We can always find vectors $\bm{s}_{A,B,C}$ and $a,b,c$ values such that $\bm{S}^2 < 1$. It is particularly clear the case $\bm{S}=\bm{0}$ as providing the maximum violation. An explicit example will be discussed below.

\section{Quasi-distributions} 

In this section, we present the most representative examples of joint distributions for a pair of observables, highlighting some of their most fundamental properties.

\subsection{Kirkwood-Dirac} 

The Kirkwood-Dirac distribution is defined as:
\begin{equation}
\label{hK}
 K_\rho (a,b)= \langle a |\rho| b \rangle\langle b |a \rangle = \mathrm{tr} \left [ \rho \hat{K} (a,b) \right ] ,
\end{equation}
where 
\begin{equation}
    \hat{K} (a,b) = |b\rangle \langle b|a\rangle \langle a | .
\end{equation}
Using relations (\ref{eigen}) and well-known properties of Pauli matrices  $\hat{K} (a,b)$ becomes:
\begin{eqnarray}
\label{hK}
&   \hat{K} (a,b)= \frac{1}{4} \left [ \left ( 1+ ab \bm{s}_A \cdot \bm{s}_B  \right ) \hat\sigma_0 \right .& \nonumber \\ & & \nonumber \\
&\left .  + \left ( a \bm{s}_A + b \bm{s}_B - i ab  \bm{s}_A \times \bm{s}_B \right )\cdot \hat{\bm{\sigma}}  \right ] .& 
\end{eqnarray}
Let us show that the four operators $\hat{K} (a,b)$ form an operator basis. To show this we note that 
\begin{equation}
    {\rm tr} \left [ \hat{K} (a,b)\hat{K}^\dagger (a^\prime,b^\prime)\right  ] = \left | \langle a | b \rangle \right |^2 \delta_{a,a^\prime} \delta_{b,b^\prime} ,
\end{equation}
where 
\begin{equation}
   \left | \langle a | b \rangle \right |^2 = \mathrm{tr} \left [ \hat{K} (a,b) \right ] = \frac{1}{2} \left ( 1+ a b \bm{s}_A \cdot \bm{s}_B \right ). 
\end{equation}
Moreover, whenever $\bm{s}_A \neq \pm \bm{s}_B$ we have ${\rm tr} \left [ O \hat{K} (a,b)\right ]=0$ for all $a,b$ if and only if $O=0$, so that the relation between observables $O$ and functions $ K_O (a,b)$ can be inverted to express $O$ in terms of $\hat{K}^\dagger (a,b)$ as
\begin{equation}
\label{OK}
   O = \sum_{a,b} \frac{1}{\left | \langle a | b \rangle \right |^2} K_O (a,b) \hat{K}^\dagger (a,b) .
\end{equation}

\bigskip

Then, the Kirkwood-Dirac distribution for an arbitrary state of the form (\ref{GSp}) is
\begin{eqnarray}
\label{Kab}
&   K_\rho(a,b)= \frac{1}{4} \left [ 1+ ab \bm{s}_A \cdot \bm{s}_B   \right .& \nonumber \\ & & \nonumber \\
&\left .  + \left ( a \bm{s}_A + b \bm{s}_B - i ab  \bm{s}_A \times \bm{s}_B \right )\cdot \bm{s} \right ] & ,
\end{eqnarray}
which in general is complex. It is interesting to compute the $AB$ correlation in the form 
\begin{equation}
\label{<AB>}
\langle AB \rangle_{K,\rho} = \sum_{a,b} a b K_\rho (a,b) ,
\end{equation}
leading to
\begin{equation}
\langle AB \rangle_{K,\rho} = \bm{s}_A \cdot \bm{s}_B  - i \left ( \bm{s}_A \times \bm{s}_B \right ) \cdot \bm{s} ,
\end{equation}
which, perhaps surprisingly, has a component independent of the system state whenever $\bm{s}_A$ and $\bm{s}_B$ are not orthogonal. This exactly reproduces the correlation of the $\hat{B}$,  $\hat{A}$  operators in Eqs. (\ref{oqc}) and (\ref{mqc}). In fact $\langle AB \rangle_{K,\rho}=\langle \hat{B} \hat{A} \rangle$,  while $ K^\ast_\rho(a,b)$ reproduces the correlation in the opposite operator ordering $\langle AB \rangle_{K^\ast,\rho}=\langle \hat{A} \hat{B} \rangle$. Interestingly $ K_\rho (a,b)$ is consistent with the case $A=B$ since in such a case
\begin{equation}
K_{\rho} (a,a^\prime) = \langle a |\rho |a \rangle \delta_{a,a^\prime}.
\end{equation}

\subsection{Margenau-Hill distribution} 

Since $ K_\rho (a,b)$ may take complex values it is not suited to test the inequality (\ref{ine}). To this end we may better consider its real part, known as the Margenau-Hill distribution
\begin{equation}
   M_\rho (a,b)= \frac{1}{2} \left [ K_\rho (a,b) +  K^\ast_\rho (a,b) \right ],
\end{equation}
this is 
\begin{equation}
\label{Mab}
 M_\rho (a,b)= \frac{1}{4} \left [ 1+ ab \bm{s}_A \cdot \bm{s}_B  + \left ( a \bm{s}_A + b \bm{s}_B \right )\cdot \bm{s} \right ]  , 
\end{equation}
and the correlation between $\bm{s}_A$ and $\bm{s}_B$ becomes 
\begin{equation}
\langle AB \rangle_{M,\rho} = \bm{s}_A \cdot \bm{s}_B  .
\end{equation}

\subsection{Wigner function} 

We may as well consider the discrete version of the well-know Wigner function $W(a,b)$ as introduced in Ref.~\cite{LP98,AL04,REK04}, that can be easily adapted to two arbitrary variables $A$ and $B$ in the form 
\begin{equation}
    W_\rho (a,b) = \mathrm{tr} \left [ \rho \hat{W} (a,b) \right ] ,
\end{equation}
where
\begin{eqnarray}
\label{hW}
&   \hat{W} (a,b)= \frac{1}{4} \left [ \left ( 1+ ab \bm{s}_A \cdot \bm{s}_B  \right ) \hat\sigma_0 \right .& \nonumber \\ & & \nonumber \\
&\left .  + \left ( a \bm{s}_A + b \bm{s}_B +ab  \bm{s}_{AB} \right )\cdot \hat{\bm{\sigma}}  \right ] &  .
\end{eqnarray}
Referring to the Kirkwood-Dirac distribution a suitable choice for the vector $\bm{s}_{AB}$ might be $\bm{s}_A \times \bm{s}_B$. However this would introduce an undesirable ordering ambiguity. So we  will endow vector $\bm{s}_{AB}$ with the desirable properties of the vector product without having to put up with the undesirable ones. These desirable properties are
\begin{eqnarray}
\label{sAB}
    & \bm{s}_{AB} = \bm{s}_{BA}, \quad \bm{s}_{AB} \cdot \bm{s}_A = \bm{s}_{AB} \cdot \bm{s}_B = 0 ,\nonumber \\
    & & \\
    & \bm{s}_{AB}^2 = 1 - \left ( \bm{s}_A \cdot \bm{s}_B\right )^2 . & \nonumber 
\end{eqnarray}

We are not attempting to define a unique Wigner function once and for all, but a representative possessing the structural properties required for the present analysis. These are here reality, proper marginals  and the following property for the kernel  $\hat{W} (a,b)$ which is common to all Wigner-like formulations, that is 
\begin{equation}
    {\rm tr} \left [ \hat{W} (a,b)\hat{W} (a^\prime,b^\prime)\right  ] = \left | \langle a | b \rangle \right |^2 \delta_{a,a^\prime} \delta_{b,b^\prime} .
\end{equation}
Since this relation implies the linear independence of the four $\hat{W}(a,b)$ operators it is  also possible here to express any operator $O$ in terms of the operators  $\hat{W} (a,b)$
\begin{equation}
\label{OW}
   O = \sum_{a,b} \frac{1}{\left | \langle a | b \rangle \right |^2}W_O (a,b) \hat{W} (a,b) .
\end{equation}

\bigskip

Then, the Wigner function for an arbitrary state of the form (\ref{GSp}) is
\begin{eqnarray}
\label{Wab}
&   W_\rho (a,b)= \frac{1}{4} \left [ 1+ ab \bm{s}_A \cdot \bm{s}_B   \right .& \nonumber \\ & & \nonumber \\
&\left .  + \left ( a \bm{s}_A + b \bm{s}_B + ab  \bm{s}_A \times \bm{s}_B \right )\cdot \bm{s} \right ] . & 
\end{eqnarray}
This is in full agreement with its relation with the Kirkwood-Dirac distribution addressed in Ref. \cite{LB26}. Here again we can compute the $AB$ correlation leading to 
\begin{equation}
\langle AB \rangle_{W,\rho} = \bm{s}_A \cdot \bm{s}_B  + \bm{s}_{AB} \cdot \bm{s} ,
\end{equation}
which differs from the symmetric-order correlation $\langle \hat{A} \hat{B} \rangle_s$ (\ref{soc}) due to the correlation term. We have also the consistency with the case $A=B$ 
\begin{equation}
W_{\rho} (a,a^\prime) = \langle a |\rho |a \rangle \delta_{a,a^\prime}.
\end{equation}

\subsection{Q/P functions} 

Let us first address a suitable candidate of a discrete Husimi Q function for this two-dimensional Hilbert space. In the standard harmonic-oscillator context Q is defined by projection of the system state on Glauber-Sudarshan coherent states. Moreover, the Q function is the statistics of a noisy simultaneous measurement of two conjugate field quadratures by double homodyne detection.  

In a two-dimensional Hilbert space all states are SU(2) coherent states, so essentially we have only to select four of them suitably related to two $A$ and $B$ observables. For definiteness we will assume $A,B$ characterized by two orthogonal unit vectors $\bm{s}_{A}\cdot \bm{s}_{B}=0$. We address an unsharp joint probability distribution for $A,B$ via the POVM
\begin{equation}
    \hat{Q} (a,b) = \frac{1}{4}  \left [ \sigma_0 + \bm{q} (a,b) \cdot \bm{\sigma} \right ] ,
\end{equation}
where
\begin{equation}
   \bm{q} (a,b) = a \gamma_A \bm{s}_A + b \gamma_B \bm{s}_B + ab \gamma_{AB} \bm{s}_{AB},
\end{equation}
with the same properties for $\bm{s}_{AB}$ already considered above in Eq. (\ref{sAB}). The $\gamma$ factors express the unsharp character of the distribution. The necessary positivity condition $\hat{Q} (a,b) \geq 0$ is granted provided that 
\begin{equation}
\label{gb}
     \gamma_A^2 + \gamma_B^2 + \gamma_{AB}^2 \leq 1 .
\end{equation}
In particular we will assume $\gamma_A^2 + \gamma_B^2 + \gamma_{AB}^2 = 1$ so that $ \hat{Q}(a,b)$ is the orthogonal projector on four pure states. In the case that all $\gamma$ were equal we will have $\gamma_A = \gamma_B = \gamma_{AB} = 1/\sqrt{3}$. Then our version of Q function is
\begin{equation}
   Q_\rho (a,b)= \mathrm{tr} \left [ \rho \hat{Q} (a,b) \right ] =
   \frac{1}{4} \left [ 1 + \bm{q} (a,b) \cdot \bm{s} \right ] .
\end{equation}
Here again we can compute the $AB$ correlation leading to 
\begin{equation}
\langle AB \rangle_{Q,\rho} = \gamma_{AB} \bm{s}_{AB} \cdot \bm{s} ,
\end{equation}
while there is no consistency with the case $A=B$ 
\begin{equation}
Q_{\rho} (a,a^\prime) \neq \langle a |\rho |a \rangle \delta_{a,a^\prime}.
\end{equation}

\bigskip

Regarding a finite dimensional P function we may follow the inversion procedure presented in Ref. \cite{SAA25}. This process leads to an operator kernel of the form 
\begin{equation}
    \hat{P} (a,b) = \frac{1}{4}  \left [ \sigma_0 + \bm{p} (a,b) \cdot \bm{\sigma} \right ] 
\end{equation}
where
\begin{equation}
   \bm{p} (a,b) = \frac{a}{\gamma_A} \bm{s}_A + \frac{b}{\gamma_B} \bm{s}_B + \frac{ab}{\gamma_{AB}} \bm{s}_{AB} ,
\end{equation}
for the same $\bm{s}_{AB}$ so we have the interesting property 
\begin{equation}
    \mathrm{tr} \left [ \hat{P} (a,b) \hat{Q} (a^\prime,b^\prime) \right ] = \frac{1}{2}  \delta_{a,a^\prime} \delta_{b,b^\prime}
\end{equation}

This defines a version of the P function as 
\begin{equation}
    P_\rho (a,b) = \mathrm{tr} \left [ \rho \hat{P} (a,b) \right ] ,
\end{equation}
that is 
\begin{equation}
   P_\rho (a,b)= \frac{1}{4} \left [ 1 + \bm{p}(a,b)\cdot \bm{s} \right ] ,
\end{equation}
and the following expansions for arbitrary operators $O$
\begin{equation}
\label{OQP}
   O = 2 \sum_{a,b} Q_O (a,b) \hat{P} (a,b) = 2 \sum_{a,b} P_O (a,b) \hat{Q} (a,b) .
\end{equation}

Here again we can compute the $AB$ correlation leading to 
\begin{equation}
\langle AB \rangle_{P,\rho} = \frac{1}{\gamma_{AB}} \bm{s}_{AB} \cdot \bm{s} ,
\end{equation}
while there is no consistency with the case $A=B$ 
\begin{equation}
P_{\rho} (a,a^\prime) \neq \langle a |\rho |a \rangle \delta_{a,a^\prime}.
\end{equation}

\subsection{Square-root distribution} 

Let us translate to the quantum domain a distribution introduced in the field of classical optics, more specifically dealing with radiometry. In radiometry complementarity forbids to talk about the light intensity at a given point propagating in a given direction. In this regard the distribution proposed in Refs. \cite{MM84,MM86} addresses a joint position-direction distribution which is nonnegative with the correct position and direction reduced intensity distributions. Such distribution is defined as 
\begin{equation}
    H(a,b) = \left | \langle a |\sqrt{\rho} | b \rangle \right |^2 ,
\end{equation}
with the characteristic of lacking linearity on $\rho/\Gamma$. Clearly it is always nonnegative and has the correct exact marginals
\begin{equation}
    \sum_b H(a,b) = \langle a |\rho| a\rangle, \quad \sum_a H(a,b) = \langle b |\rho| b\rangle. 
\end{equation}
This can be considered as the quantum analogue of the radiometric distribution introduced by Mart\'{i}nez-Herrero and Mej\'{i}as,  \cite{MM84,MM86}.

\bigskip

For pure states $\rho = |\psi \rangle \langle \psi |$ the expression is very simple since  $\sqrt{\rho} = |\psi \rangle \langle \psi |$ and then
\begin{equation}
\label{fac}
    H(a,b) = p_A(a)p_B(b) ,
\end{equation}
where $p_X(x) = |\langle x |\psi \rangle |^2$ are the corresponding probabilities $X=A,B$, $x=a,b$. For mixed states and for the sake of dealing easily with the square root let us begin from a slightly modified version of Eq. (\ref{GSp}) in the form
\begin{equation}
    \rho = \Gamma = \frac{1}{2} \left ( \hat\sigma_0+ \mu \bm{s}\cdot \bm{\sigma} \right ) ,
\end{equation}
where $\bm{s}$ is a unit vector $\bm{s}^2=1$ and $\mu$ is a real number $1 \geq \mu \geq0$ expressing the purity of $\rho$, that is that the state is pure if $\mu=1$. Then 
\begin{equation}
\rho = \frac{1+\mu}{2} |+\rangle \langle + | +  \frac{1-\mu}{2} |-\rangle \langle - | ,
\end{equation}
where $|\pm \rangle$ are the eigenstates of $\bm{s}\cdot \bm{\sigma}$ with eigenvalues $\pm 1$ respectively, so that 
\begin{equation}
|\pm \rangle \langle \pm | = \frac{1}{2} \left ( \sigma_0  \pm \bm{s}\cdot \bm{\sigma}\right ) .
\end{equation}
Then
\begin{equation}
    \sqrt{\rho} = \frac{1}{2} \left (\eta_+ \sigma_0 + \eta_- \bm{s}\cdot \bm{\sigma} \right )  ,
\end{equation}
where 
\begin{equation}
\eta_{\pm} = \sqrt{\frac{1+\mu}{2}} \pm \sqrt{\frac{1-\mu}{2}} .
\end{equation}
This leads to the following final form for the distribution
\begin{eqnarray}
    & H(a,b) = \frac{1}{4} \left [ 1 + ab\sqrt{1-\mu^2} \bm{s}_B \cdot \bm{s}_A \right . & \nonumber \\
    & & \nonumber \\
    & + a \mu \bm{s}\cdot \bm{s}_A + b \mu \bm{s}\cdot \bm{s}_B  & \nonumber \\
     & & \nonumber \\
    &\left . + ab\left (1- \sqrt{1-\mu^2} \right ) (\bm{s}\cdot \bm{s}_A)(\bm{s}\cdot \bm{s}_B ) \right ] . &  
\end{eqnarray}
Here again we can compute the $AB$ correlation with both state-dependent and state-independent contributions
\begin{eqnarray}
& \langle AB \rangle_{H,\rho} = \sqrt{1-\mu^2} \bm{s}_B \cdot \bm{s}_A & \nonumber \\
& & \nonumber \\ 
& +\left (1- \sqrt{1-\mu^2} \right ) (\bm{s}\cdot \bm{s}_A)(\bm{s}\cdot \bm{s}_B ) , & 
\end{eqnarray}
that for pure states $\mu=1$ is just $\langle AB \rangle_{H,\rho} = \langle A \rangle \langle B \rangle$. Finally there seems to be no $A=B$ consistency.

\section{Inequality test of quasi-distributions} 

Next we examine whether the joint distributions we have just defined satisfy the inequality  (\ref{ine}), and if so, under what conditions.  

\subsection{Margenau-Hill distribution} 

Let us begin with the Margenau-Hill distribution that follows the general law (\ref{gf}) 
\begin{equation}
    M_{AC} (a,c)+M_{BC}(b,-c) = M_{AB}(a,b)+ \eta ,
\end{equation}
with 
\begin{equation}
\label{eta}
    \eta = \frac{1}{4} \left ( 1 - ab \bm{s}_A \cdot \bm{s}_B + ac \bm{s}_A \cdot \bm{s}_C - bc \bm{s}_B \cdot \bm{s}_C \right )  ,
\end{equation}
which can be as well expressed as in Eq. (\ref{eqc})
\begin{equation}
\label{eM}
    \eta = \frac{1}{8} \left ( \bm{S}_M^2 -1\right ) ,  
\end{equation}
with 
\begin{equation}
\label{MS}
    \bm{S}_M =a \bm{s}_A - b \bm{s}_B + c \bm{s}_C 
\end{equation}
The fulfillment of the inequality depends on the observables considered $A,B,C$ and the values $a,b,c$, but it does not depend on the state $\rho$. So, the fulfillment of the inequality for $\eta \geq 0$ holds even if the distributions involved take negative values, and vice-versa, it can be violated by for states whose $M_{XY} (x,y)$ distribution are legitimate probabilities. 

 For example, if the observables are such that $\bm{s}_{A,B,C}$ are mutually orthogonal then $\eta =1/4$  and the inequality is always satisfied. On the other hand, we can present a clear case of violation by considering the case in which $\bm{s}_A$, $\bm{s}_B$ and $\bm{s}_C$ lie in the same plane, with $\bm{s}_A$ and $\bm{s}_C$ forming an angle of $2\pi/3$, while $\bm{s}_B$ lies between them forming an angle of $\pi/3$ with both of them, so that
\begin{equation}
\label{vMH}
    \bm{s}_A \cdot \bm{s}_B = \bm{s}_B \cdot \bm{s}_C = \frac{1}{2}, \quad \bm{s}_A \cdot \bm{s}_C = -\frac{1}{2} .
\end{equation}
Then, for example for $a=b=c =1$ we have $\eta = -1/8$ and the inequality is violated. Actually this is the minimum value possible for $\eta$ since after Eq. (\ref{eM}) $\eta \geq -1/8$, the equality being reached when the vector $\bm{S}_M$ in Eq. (\ref{MS}) vanishes $\bm{S}_M=\bm{0}$.

\subsection{Wigner function}

The Wigner function also follows the general law (\ref{gf}) 
\begin{equation}
    W_{AC}(a,c)+W_{BC}(b,-c) = W_{AB} (a,b)+ \eta ,
\end{equation}
where in $\eta$ we can distinguish two contributions $\eta= \eta_\mathrm{si}+\eta_\mathrm{sd}$, where 
$\eta_\mathrm{si}$ is the same state-independent contribution in Eq. (\ref{eta}), while $\eta_\mathrm{sd}$ is an state-dependent term with 
\begin{equation}
\label{eta2}
\eta_\mathrm{sd} = \frac{1}{4} \bm{S}_W \cdot \bm{s}  ,
\end{equation}
where 
\begin{equation}
\label{SW}
 \bm{S}_W= -ab \bm{s}_{AB}  + ac \bm{s}_{AC}  - bc \bm{s}_{BC} .
\end{equation}

We can offer two simple and interesting cases of inequality violation. For example we may consider the maximally mixed state $\rho=\sigma_0/2$, for which $\bm{s}=\bm{0}$ and then $\eta_\mathrm{sd}=0$. Then we are exactly in the same conditions of the Margenau-Hill distribution, with a clear violation of the inequality for the choice of observables in Eq. (\ref{vMH}). It is worth noting that this violation occurs for a state for which $W(a,b)$ is nonnegative for all pairs of observables. 

Moreover, we can present the case of mutually orthogonal $\bm{s}_X$ which implies mutually orthogonal $\bm{s}_{XY}$, $X,Y=A,B,C$, so that $ \eta_\mathrm{si} =1/4$ and
\begin{equation}
   \eta= \frac{1}{4} \left ( 1 + \bm{S}_W \cdot \bm{s} \right ), \quad |\bm{S}_W |= \sqrt{3} .
\end{equation}
The inequality is then violated for states with $\bm{S}_W \cdot \bm{s} \leq -1$, the minimum for $\eta$ arising for $\bm{s} = - \bm{S}_W /|\bm{S}_W |$.

\subsection{Q/P functions} 

Here again in accordance with \ref{gf}) we have 
\begin{equation}
    Q_{AC}(a,c)+Q_{BC}(b,-c) = Q_{AB}(a,b) + \eta ,
\end{equation}
where
\begin{equation}
\eta = \frac{1}{4} \left ( 1 + \bm{S}_Q \cdot \bm{s} \right ) ,
\end{equation}
being 
\begin{equation}
\label{Sq}
 \bm{S}_Q=   -ab \gamma_{AB} \bm{s}_{AB}  + ac \gamma_{AC}\bm{s}_{AC}  - bc \gamma_{BC}\bm{s}_{BC} .
\end{equation}
In this case, given that the vectors $\bm{s}_X$ are orthogonal so they are the $\bm{s}_{XY}$, $X,Y=A,B,C$, and given that the $\gamma$ factors are so bounded according to Eq. (\ref{gb}) we have $|\bm{S}_Q| <1$ and no state can violate the inequality. 

\bigskip

On the other hand the P function may easily violate the inequality since, again after the general law (\ref{gf})
\begin{equation}
    P_{AC}(a,c)+P_{BC}(b,-c) = P_{AB}(a,b) + \eta,
\end{equation}
where 
\begin{equation}
 \eta = \frac{1}{4} \left ( 1 + \bm{S}_P \cdot \bm{s} \right ) 
\end{equation}
with 
\begin{equation}
\label{Sp}
 \bm{S}_P =   -\frac{ab}{\gamma_{AB}} \bm{s}_{AB}  + \frac{ac}{\gamma_{AC}}\bm{s}_{AC}  - \frac{bc}{\gamma_{BC}}\bm{s}_{BC} ,
\end{equation}
so that in the same conditions of the Q function regarding the $\bm{s}_X$ and $\bm{s}_{XY}$ vectors, we have in this case $|\bm{S}_P| >1$ and the inequality can be violated for every $A,B,C$ by a suitably chosen system state $\bm{s}$.

\subsection{Square-root distribution} 

For all pure states it is easy to see that the inequality is always satisfied for all observables. This is because the factorization (\ref{fac}) leads us to express the inequality as
\begin{equation}
\label{fi}
    p_A (a)p_B (b) \leq  p_A (a)p_C (c)+ p_B (b) p_C (-c).
\end{equation}
To show this we can consider that for the right-hand side 
\begin{equation}
    p_A (a)x+(1-x)p_B (b) \geq {\rm min} \left [ p_A (a),p_B (b) \right ] ,
\end{equation}
$1\geq x \geq 0$ where $x=p_C(c)$, while for the left-hand side 
\begin{equation}
    p_A (a)p_B (b) \leq  {\rm min} \left [ p_A (a),p_B (b) \right ] ,
\end{equation}
as far as $p_A (a)p_B (b) \leq p_A (a)$ and $p_A (a)p_B (b) \leq p_B (b)$. Therefore
\begin{equation}
    p_A (a)p_B (b) \leq  {\rm min} \left [ p_A (a),p_B (b) \right ] \leq  p_A (a)x+(1-x)p_B (b) ,
\end{equation}
and the inequality (\ref{fi}) always holds. 

\bigskip

It is also rather simple to examine the case where $\bm{s}_{A,B,C}$ are mutually orthogonal, say along axes $X$, $Y$ and $Z$, respectively, so that 
\begin{equation}
    H_{AC}(a,c)+H_{BC}(b,-c) = H_{AB}(a,b)+ \eta ,
\end{equation}
where 
\begin{equation}
    \eta = \frac{1}{4} \left [ 1 + \left (1- \sqrt{1-\mu^2} \right ) \left ( -ab s_x s_z + ac s_x s_y - bc s_y s_z \right ) \right ].
\end{equation}
the minimum of $-ab s_x s_z + ac s_x s_y - bc s_y s_z $ when $\bm{s}$ and $a,b,c$ are varied is $-1/2$, as it can be seen expanding $(as_x+cs_y-bs_z)^2$. Then 
\begin{equation}
\eta \geq \frac{1}{8} \left ( 1 +\sqrt{1-\mu^2} \right ) \geq \frac{1}{8},
\end{equation}
and the equality is always satisfied. 

\bigskip

The analysis of the general case is rather involved. We can show that the inequality can be violated again for a maximally mixed state, that is $\mu =0$ and $\rho = \sigma_0/2$. In such a case,
\begin{equation}
     H_{XY} (x,y) = \frac{1}{4} \left ( 1+xy \bm{s}_X \cdot \bm{s}_Y  \right )
\end{equation}
so that $\eta$ is exactly the same in Eq. (\ref{eta}), so there is a clear violation for the same choice of observables and values considered there.

\section{Noisy version of the inequality} 

Another way to obtain joint distributions for pairs of arbitrary operators is to perform simultaneous measurements that, although noisy, contain complete information about the observables in question. This may be in particular the case of the Q function considered above. In particular this may provide valid practical estimates for the correlations in the version (\ref{eta}) of the inequality (\ref{ine}). 

To this end we can begin with the noisy version of (\ref{nvpXY}) 
\begin{equation}
    \tilde{p}_{XY} (x,y) = \frac{1}{4} \left ( 1 + x \langle X \rangle_{\tilde{p}} + y \langle Y \rangle_{\tilde{p}}+ xy \langle XY \rangle_{\tilde{p}} \right ) .
\end{equation}
Typically there is a clear relation between the noisy $\langle X \rangle_{\tilde{p}}$ and exact $\langle X \rangle_p$ values  
\begin{equation}
    \langle X \rangle_{\tilde{p}} = \gamma_X \langle X \rangle_p ,
\end{equation}
for $X,Y=A,B,C$, where $\gamma$ are the noise factors $1 \geq \gamma \geq 0$. This certainly means an increase of noise since for dichotomic variables the uncertainty depends just on the mean values
\begin{equation}
    \Delta^2 X = 1 - \langle X \rangle ,
\end{equation}
so that a smaller mean value implies larger uncertainty. Then we can obtain the exact mean values as 
$\langle X \rangle_p = \langle X \rangle_{\tilde{p}} /\gamma_X$ and likewise for the correlation term
\begin{equation}
    \langle X Y\rangle_p = \frac{\langle X Y\rangle_{\tilde{p}}}{\gamma_X \gamma_Y} ,
\end{equation}
so we have already a practical version of the $\eta$ factor in Eqs. (\ref{gf}), (\ref{eta})
\begin{equation}
\eta = \frac{1}{4} \left ( 1 - ab \frac{\langle AB\rangle_{\tilde{p}}}{\gamma_A \gamma_B} +ac \frac{\langle AC \rangle_{\tilde{p}}}{\gamma_A \gamma_C}- cb \frac{\langle C B \rangle_{\tilde{p}}}{\gamma_C \gamma_B} \right ).
\end{equation}
To proceed further let us assume a typical form for the correlations $\langle X Y\rangle_{\tilde{p}}$, that is for example 
\begin{equation}
\langle X Y\rangle_{\tilde{p}} = \gamma_{XY} \bm{s}_{XY} \cdot \bm{s} ,
\end{equation}
so that $\eta$ becomes
\begin{equation}
\eta = \frac{1}{4} \left ( 1 + \bm{S}_N \cdot \bm{s} \right ) ,
\end{equation}
where 
\begin{equation}
\label{Sq}
 \bm{S}_N =   -ab \frac{\gamma_{AB}}{\gamma_A \gamma_B} \bm{s}_{AB}  + ac \frac{\gamma_{AC}}{\gamma_A \gamma_C}\bm{s}_{AC}  - bc \frac{\gamma_{BC}}{\gamma_B \gamma_C}\bm{s}_{BC} .
\end{equation}

There are very simple situations showing that the bound $\eta \geq 0$ can be beaten. This is for example the case of mutually orthogonal $\bm{s}_{XY}$ and $\gamma_X=\gamma_Y = \gamma_{XY}= \gamma$, so that $|\bm{S}_N| = \sqrt{3}/\gamma$. The minimum $\eta$ holds when $\bm{s}=-\bm{S}_N/|\bm{S}_N|$ leading to 
\begin{equation} 
\eta_\mathrm{min} = \frac{1}{4} \left ( 1- \frac{\sqrt{3}}{\gamma} \right ) ,
\end{equation}
so that $\eta_\mathrm{min} <0$ for all $\gamma$.

\bigskip

\section{Conclusions} 

We have derived a family of statistical inequalities whose fulfillment is equivalent to the existence of an underlying genuine joint probability distribution. These inequalities are based on correlations between observables that are generally incompatible. This is perhaps one of the most interesting points, in that quantum theory does not uniquely prescribe such correlations due to the well-known problem of the order of operators that do not commute. 

With them we have tested the behavior of several representative quantum quasidistributions as well as joint distributions obtained from noisy simultaneous measurements. Particularly remarkable are the violations obtained for the maximally mixed state and, in some cases, for every quantum state.

These results indicate that the observed nonclassicality cannot always be regarded as a property of the quantum state alone. Rather, it reflects structural features of the statistical framework used to represent incompatible observables, emphasizing the fundamental role played by the impossibility of embedding quantum statistics into a single classical probability space.

\end{document}